\documentclass[aps,prb,twocolumn,groupedaddress,showpacs]{revtex4}

\usepackage{graphicx}
\usepackage{dcolumn}
\usepackage{bm}

\begin{document}

\title{
Key role of hybridization between actinide $5f$ and oxygen $2p$ orbitals\\
for electronic structure of actinide dioxides
}

\author{Yu Hasegawa$^1$, Takahiro Maehira$^2$, and Takashi Hotta$^{1,3}$}

\affiliation{$^1$Department of Physics, Tokyo Metropolitan University,
Hachioji, Tokyo 192-0397, Japan \\
$^2$Faculty of Science, University of The Ryukyus, Nishihara,
Okinawa 903-0213, Japan \\
$^3$Advanced Science Research Center,
Japan Atomic Energy Agency, Tokai, Ibaraki 319-1195, Japan}

\date{\today}

\begin{abstract}
In order to promote our understanding on electronic structure of
actinide dioxides, we construct a tight-binding model composed of
actinide $5f$ and oxygen $2p$ electrons,
which is called $f$-$p$ model.
After the diagonalization of the $f$-$p$ model,
we compare the eigenenergies in the first Brillouin zone
with the results of relativistic band-structure calculations.
Here we emphasize a key role of $f$-$p$ hybridization
in order to understand the electronic structure of actinide dioxides.
In particular, it is found that
the position of energy levels of $\Gamma_7$ and $\Gamma_8$ states
determined from crystalline electric field potentials
depends on the $f$-$p$ hybridization.
We clarify the condition on the $f$-$p$ hybridization to explain
the electronic structure which is consistent with the local 
crystalline electric field state.
We briefly discuss the region of the absolute values of the Slater-Koster
integrals for $f$-$p$ hybridization concerning the appearance of
octupole ordering in NpO$_2$.
\end{abstract}



\maketitle

\section{Introduction}

Actinide dioxides form a group of important materials from
technological viewpoints of a nuclear reactor fuel and
a heterogeneous catalyst.
On the other hand, this material group has been actively
investigated also from a viewpoint of basic science
because of its high symmetry of the fluorite structure of
the space group $Fm3m$.\cite{review1,Hotta-review,review2}
In the circumstance of such high symmetry of crystal structure,
it is possible to observe peculiar ordering of multipole higher than
dipole, when we change the kind of actinide ions.
Among several magnetic properties of actinide dioxides,
a mysterious low-temperature ordered phase of NpO$_2$ has
attracted continuous attention in the research field of
condensed matter physics.

The phase transition in NpO$_2$ has been confirmed in 1953
from the observation of a peak in the specific heat
around 25 K.\cite{Westrum}
Due to the behavior of magnetic susceptibility,\cite{Ross,Erdos}
it has been first considered that the antiferromagnetic order occurs.
Unfortunately, no dipole moments have been observed in the low-temperature
phase and a puzzling situation has continued.
Neutron scattering experiments have revealed that the
ground state of the crystalline electric field (CEF)
potential is $\Gamma_8$,
\cite{Fournier} which carries multipole moments.
Then, from several phenomenological works on the ordered phase,
a key role of octupole degree of freedom has been focused.
\cite{Santini,Paixao,Caciuffo,Lovesey,Kiss}
In fact, the octupole ordering has been strongly suggested
by $^{17}$O-NMR experiment \cite{Tokunaga}
and also by inelastic neutron scattering study.\cite{Magnani}
As for the microscopic origin of such higher-order multipole ordering,
it has been shown that octupole order is stabilized by
the orbital-dependent superexchange interaction,
obtained by the second-order perturbation of $5f$ electron
hopping in the $\Gamma_8$ degenerate Hubbard model
on a fcc lattice.\cite{Kubo1,Kubo2,Kubo3}
Recently, significant contribution of dotriacontapole moment
has been also pointed out.\cite{Santini2,Suzuki}

Since the multipole moments originate from $5f$ electrons,
it seems to be natural to consider the Hubbard-like model of
$5f$ electrons.
However, from the crystal structure of actinide dioxides,
it is also important to include explicitly $2p$ electrons,
since actinide ion is surrounded by eight oxygens and
the main hopping process between nearest neighbor sites
should occur from the $f$-$p$ hybridization.
In this sense, $f$-$p$ model is more realistic Hamiltonian
for actinide dioxides.
In fact, the $f$-$p$ model for actinide dioxides has been analyzed
in the fourth-order perturbation theory in terms of $f$-$p$ hybridization.
\cite{Kubo4}
Then, it has been revealed that octupole order actually occurs
even when we include oxygen $2p$ electrons.
However, there has been a peculiar point that
the octupole phase appears {\it only} for the small absolute value of
$(fp\pi)/(fp\sigma)$, where $(fp\sigma)$ and $(fp\pi)$
are Slater-Koster integrals between $f$ and $p$ orbitals.
The reason of the sensitivity of the octupole ordered phase concerning
the $f$-$p$ hybridization has not been understood yet.

In order to clarify the role of $f$-$p$ hybridization
for the appearance of octupole ordering,
Maehira and Hotta have performed the band-structure
calculations for actinide dioxides
by a relativistic linear augmented-plane-wave method
with the exchange-correlation potential
in a local density approximation.\cite{Maehira}
It has been found that the energy bands in the vicinity of
the Fermi level are mainly due to the hybridization between
actinide $5f$ and oxygen $2p$ electrons.
It has been also pointed out that the electronic structure
at the $\Gamma$ point in the first Brillouin zone
is not consistent with that of the local CEF state.
One reason for this inconsistency is that
the CEF potentials are not satisfactorily included
in the calculations,
but it is difficult to control the magnitudes of CEF potential
and $f$-$p$ hybridization
in the band-structure calculations.
It is highly requested to reveal the role of
$f$-$p$ hybridization for the simultaneous explanation
of the octupole ordering and the local CEF states.

In this paper, in order to clarify the roles of
hybridization between actinide $5f$ and oxygen $2p$ electrons
for the electronic structure of actinide dioxides,
we analyze the tight-binding $f$-$p$ model in detail.
Except for the Slater-Koster integrals of $(fp\pi)$ and $(fp\sigma)$,
we determine the parameters in the model from the comparison
with experimental results and band-structure calculations.
In order to reproduce the result of the relativistic band-structure
calculations and obtain the electronic structure consistent with
the local CEF state, we find that the Slater-Koster parameters
for $f$-$p$ hybridization should be limited in a certain range.
A typical result is found for $(fp\pi) \approx 0$ and
$(fp\sigma) \approx 1$ eV,
which is consistent with the condition for the appearance of
the octupole ordering.

The organization of this paper is as follows.
In Sec.~II, in order to make this paper self-contained,
we briefly review the relativistic band-structure
calculations for actinide dioxides.
It is meaningful to define the problems
included in the band-structure calculations.
In Sec.~III, we explain a way to construct the $f$-$p$ model
in the tight-binding approximation.
Then, we determine the parameters of the model,
except for $(fp\sigma)$ and $(fp\pi)$, from the comparison
with the experimental and band-structure calculation results.
In Sec.~IV, we depict the energy band structure of the $f$-$p$ model
by changing the values of $f$-$p$ hybridization.
We deduce the reasonable regions for $(fp\sigma)$ and $(fp\pi)$.
In Sec.~V, we discuss some future problems
concerning the electronic structure of actinide dioxides.
Finally, we summarize this paper.
Throughout this paper, we use such units as $\hbar$=$k_{\rm B}$=1.

\section{Brief review of band-structure calculations for actinide dioxides}

Let us briefly review the band-structure calculation results
in order to clarify the problem in the understanding of
electronic structure of actinide dioxides.
As for details, readers should consult Ref.~\onlinecite{Maehira}.

We have performed the calculations by using
the relativistic linear augmented-plane-wave (RLAPW) method.
We assume that all $5f$ electrons are itinerant and
perform the calculations in the paramagnetic phase.
Note that we should take into account relativity
even in the calculations for solid state physics
because of large atomic numbers of the constituent atoms.
The spatial shape of the one-electron potential is determined
in the muffin-tin approximation.
We use the exchange and correlation potential
in a local density approximation (LDA).
The self-consistent calculation is carried out
for the experimental lattice constant for actinide dioxides.

\begin{figure}[t]
\begin{center}
\includegraphics[width=8cm]{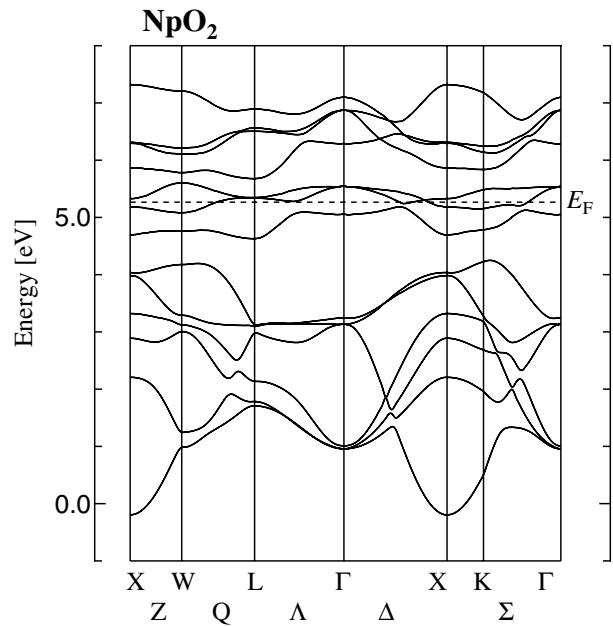}
\caption{Energy band structure of NpO$_2$ obtained by
the self-consistent RLAPW method.
Note that we pick up only $5f$ and $2p$ bands around $E_{\rm F}$,
which indicate the position of the Fermi level.}
\label{fig1}
\end{center}
\end{figure}

In Fig.~1, we show a typical result for NpO$_2$
along the symmetry axes in the Brillouin zone.
In the energy band structure in the vicinity of $E_{\rm F}$,
there always occurs a hybridization between
actinide $5f$ and oxygen $2p$ states for actinide dioxides.
The lowest six bands originate from the oxygen $2p$ states and 
are fully occupied and the width of oxygen $2p$ band is about 4.76 eV.
Narrow bands lying in the region 4.5-7.5 eV are the $5f$ bands 
which are split into two subbands by the spin-orbit interaction.
The spin-orbit splitting in the $5f$ states is
estimated as 0.95 eV, which is consistent with that for
isolated neutral Np atom.
Note that in the LDA calculation, we find the metallic state for NpO$_2$,
not the insulating state. This point will be discussed later.

Here we remark that $\Gamma_{7}$ doublet and $\Gamma_{8}$ quartet
levels appear around $E_{\rm F}$ at the $\Gamma$ point.
It should be noted that the $\Gamma_7$ level is lower than
the $\Gamma_8$ in our band-structure calculations.
However, from the CEF analysis on the basis of the $j$-$j$ coupling scheme,
$\Gamma_8$ becomes lower than $\Gamma_7$ in actinide dioxides.
When we accommodate $5f$ electrons in $\Gamma_8$ orbitals,
we obtain $\Gamma_5$ triplet for $n$=2, $\Gamma_8$ quartet for $n$=3,
and $\Gamma_1$ singlet for $n$=4, as experimentally found
in the CEF ground states of UO$_2$ \cite{Amoretti},
NpO$_2$ \cite{Fournier}, and PuO$_2$ \cite{Kern1,Kern2}.
Note here that $n$ denotes the number of local $5f$ electrons.

In order to resolve the problems,
it is necessary to improve the method to include
the effect of CEF potentials
beyond the simple estimation of the Madelung potential energy.
However, it is a difficult task to perform such improvement
concerning the formulation of the band-structure calculation.
Thus, in this paper, we choose an alternative method to
exploit the tight-binding $f$-$p$ model
for the purpose to understand the
role of $f$-$p$ hybridization for the change of CEF states
in the tight-binding model.
By changing the parameters in the $f$-$p$ model,
we attempt to clarify the key quantities which characterize
the electronic structure of actinide dioxides.

\section{Tight-binding approximation}

\subsection{Crystal structure and unit cell}

Before proceeding to the construction of
a tight-binding model for actinide dioxides
with the fluorite structure,
first let us define the unit cell including one actinide
ion and two oxygen ions,
as shown in the Fig.~2.
The basis vectors of the fcc lattice are given by 
$\bm{a}_1=(a/2,a/2,0)$,
$\bm{a}_2=(0,a/2,a/2)$,
and $\bm{a}_3=(a/2,0,a/2)$,
where $a$ is the lattice constant.
Thus, in Fig.~2, positions of adjacent four actinide ions are given by
$\bm{i}$, $\bm{i}+\bm{a}_1$, $\bm{i}+\bm{a}_2$, and $\bm{i}+\bm{a}_3$,
where $\bm{i}$ denotes the position vector for one actinide ion.

The positions of eight nearest-neighbor oxygen ions are given by
$\bm{b}_1=(a/4,a/4,a/4)$,
$\bm{b}_2=(-a/4,a/4,a/4)$,
$\bm{b}_3=(a/4,-a/4,a/4)$,
$\bm{b}_4=(-a/4,-a/4,a/4)$,
$\bm{b}_5=(-a/4,-a/4,-a/4)$,
$\bm{b}_6=(-a/4,a/4,-a/4)$,
$\bm{b}_7=(a/4,-a/4,-a/4)$,
and $\bm{b}_8=(a/4,a/4,-a/4)$.
Note that the two oxygens, O1 and O2, in the same unit cell are
specified by $\bm{b}_1$ for O1 and $\bm{b}_5$ for O2, respectively.

\begin{figure}[t]
\begin{center}
\includegraphics[width=8.5cm]{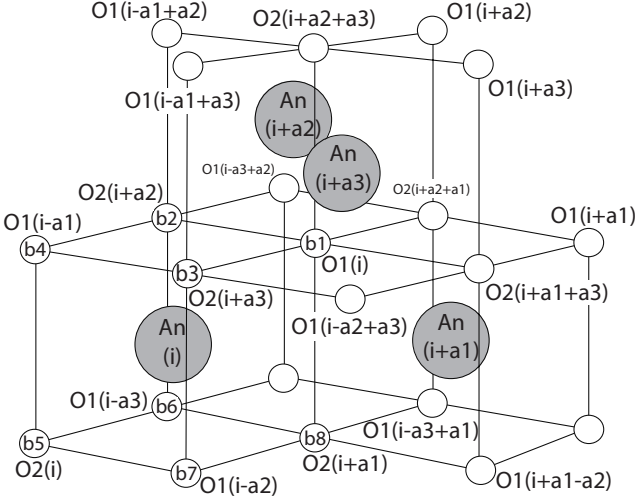}
\caption{Crystal structure of AnO$_2$.
Large solid circles denote actinide ions of which positions are given
by $\bm{i}$, $\bm{i}+\bm{a}_1$, $\bm{i}+\bm{a}_2$, and $\bm{i}+\bm{a}_3$.
Small open circles indicate oxygen ions.
Note that two oxygen ions, O1($\bm{i}$) and O2($\bm{i}$), are included
in the unit cell containing one actinide ion specified by $\bm{i}$.
}
\end{center}
\end{figure}

\subsection{CEF state}

Now we define the basis of $f$ electrons
when we consider the electronic model for
actinide dioxides with the fluorite structure.
For the purpose, we solve the problem of
one $f$ electron in the CEF potential.
The CEF Hamiltonian is written as
\begin{equation}
 H_{\rm CEF} = \sum_{\bm{i},m,m',\sigma}
  B_{m,m'} f^{\dag}_{\bm{i}m\sigma}f_{\bm{i}m\sigma},
\end{equation}
where $f_{\bm{i}m\sigma}$ is the annihilation operator
of $f$ electron at site $\bm{i}$ with spin $\sigma$
in the orbital specified by $m$.
Note that $m$ is
the $z$-component $m$ of angular momentum $\ell=3$.
We note also that the spin-orbit coupling is not
included at this stage.

Since the fluorite structure belongs to O$_{h}$ point group,
$B_{m,m'}$ is given by using a couple of CEF parameters
$B^{0}_{4}$ and $B^{0}_{6}$ for angular momentum $\ell=3$
as\cite{LLW,Hutchings}
\begin{eqnarray}
&&B_{3,3}=B_{-3,-3}=180B^{0}_{4}+180B^{0}_{6}\nonumber \\
&&B_{2,2}=B_{-2,-2}=-420B^{0}_{4}-1080B^{0}_{6}\nonumber \\
&&B_{1,1}=B_{-1,-1}=60B^{0}_{4}+2700B^{0}_{6} \\
&&B_{0,0}=360B^{0}_{4}-3600B^{0}_{6}\nonumber \\
&&B_{3,-1}=B_{-3,1}=60\sqrt[]{15}(B^{0}_{4}-21B^{0}_{6})\nonumber \\
&&B_{2,-2}=300B^{0}_{4}+7560B^{0}_{6}\nonumber,
\end{eqnarray}
Note the relation of $B_{m,m'}=B_{m',m}$.

After performing the diagonalization of $H_{\rm CEF}$,
we obtain three kinds of CEF states:
$\Gamma_2$ singlet (xyz),
$\Gamma_4$ triplet (${\rm x(5x^2-3r^2)}$, ${\rm y(5y^2-3r^2)}$,
${\rm z(5z^2-3r^2)}$),
and $\Gamma_5$ triplet (${\rm x(y^2-z^2)}$, ${\rm y(z^2-x^2)}$,
${\rm z(x^2-y^2)}$).
The corresponding CEF energies are given by
$E(\Gamma_2)=-720(B^{0}_{4}+12B^{0}_{6})$,
$E(\Gamma_4)=360(B^{0}_{4}-10B^{0}_{6})$,
and
$E(\Gamma_5)=-120(B^{0}_{4}-54B^{0}_{6})$.
Note that these seven states are elements of
cubic harmonics for $\ell=3$.
In the traditional notation, we express CEF parameters $B_4^0$ and $B_6^0$
as $B_{4}^0=Wx/F(4)$ and $B_{6}^0=W(1-|x|)/F(6)$with $F(4)$=15 and 
$F(6)$=180 for angular momentum $\ell=3$.\cite{Hutchings}
Note that $x$ specifies the CEF scheme for $O_{\rm h}$ point group,
while $W$ determines an energy scale for the CEF potential.
\begin{eqnarray}
E(\Gamma_2)=-48W[x+(1-|x|)], \nonumber \\
E(\Gamma_4)=4W[6x-5(1-|x|)], \\
E(\Gamma_5)=-4W[2x-9(1-|x|)]. \nonumber
\end{eqnarray}
The value of $W$ and $x$ will be discussed later.

Since we will construct the model in the cubic system,
it seems to be natural to use these cubic harmonics
as $f$-electron basis function.
Thus, in the following, we define $\mu$ as
the index to distinguish the orbitals of cubic
harmonics.
Note that $\mu$ takes $1 \sim 7$ and
the definitions are as follows:
1: xyz, 2:${\rm x(5x^2-3r^2)}$, 3:${\rm y(5y^2-3r^2)}$,
4: ${\rm z(5z^2-3r^2)}$, 5:${\rm x(y^2-z^2)}$,
6:${\rm y(z^2-x^2)}$, and 7:${\rm z(x^2-y^2)}$.
The corresponding energy $E_{\mu}$ is given by the above equations.

\subsection{Hamiltonian}

The Hamiltonian is given by
\begin{equation}
  H = H_{f} + H_{fp} + H_{p},
\end{equation}
where $H_{f}$ and $H_{p}$ denote
$f$- and $p$-electron part, respectively, while
$H_{fp}$ indicates $f$-$p$ hybridization term.
In the following, we explain the construction
of each term.

\subsubsection{$f$-electron term}

The $f$-electron part is given by
\begin{eqnarray}
 \label{eq:Hf}
 H_{f} &=& \sum_{\bm{k},\mu,\mu',\sigma}
 [\varepsilon^{ff}_{\bm{k}\mu\mu'}
 +(E_f+ E_{\mu})\delta_{\mu\mu'}]
 f^{\dag}_{\bm{k}\mu\sigma}f_{\bm{k}\mu'\sigma} \nonumber\\
 &+& \lambda \sum_{\bm{k},\mu,\mu',\sigma,\sigma'}
 \zeta_{\mu,\sigma,\mu',\sigma'}
 f^{\dag}_{\bm{k}\mu\sigma}f_{\bm{k}\mu'\sigma'},
\end{eqnarray}
where $f_{\bm{k}\mu\sigma}$ is the annihilation operator
of $f$ electron with spin $\sigma$ in the orbital $\mu$,
$\varepsilon^{ff}_{\bm{k}\mu\mu'}$
is the $f$-electron dispersion due to
the hopping between nearest neighbor actinide ions,
$E_{f}$ is the $f$-electron level,
$E_{\mu}$ denotes the CEF potential energy of $\mu$ orbital,
$\lambda$ is the spin-orbit interaction,
and $\zeta$ is the spin-orbit matrix element.

Concerning the expression of the spin-orbit coupling,
it is necessary to step back to the basis of the spherical harmonics.
On the basis labelled by $m$,
the spin-orbit interaction $\zeta_{m,\sigma,m',\sigma'}$
is expressed as
\begin{eqnarray}
&&\zeta_{m,\pm1/2,m,\pm1/2}=\pm m/2, \nonumber \\
&&\zeta_{m\pm 1/2,\mp 1/2, m, \pm 1/2}=\sqrt{12-m(m \pm 1)}/2,
\end{eqnarray}
and zero for the other cases.
By transforming the basis from $m$ to $\mu$,
we obtain $\zeta_{\mu,\sigma,\mu',\sigma'}$ in eq.~(\ref{eq:Hf}).

The $f$-electron dispersion in eq.~(\ref{eq:Hf}) is expressed as
\begin{equation}
 \varepsilon^{ff}_{\bm{k}\mu\mu'}
 = \sum_{\bm{a}}e^{i \bm{k} \cdot \bm{a}}
 t^{ff}_{\mu\mu'}(\bm{a}),
\end{equation}
where $\bm{a}$ denotes the vectors connecting
twelve nearest neighbor sites of the fcc lattice and
$t^{ff}_{\mu\mu'}(\bm{a})$ indicates the $f$-electron
hopping amplitude between $\mu$ and $\mu'$ orbitals
along the direction of $\bm{a}$.
Here we note that $\bm{a}$ runs among
$\pm \bm{a}_1$,
$\pm \bm{a}_2$,
$\pm \bm{a}_3$,
$\pm ( \bm{a}_2-\bm{a}_3 )$,
$\pm ( \bm{a}_3-\bm{a}_1 )$, 
and $\pm ( \bm{a}_1-\bm{a}_2 )$.
The hopping integral $t^{ff}_{\mu\mu'}(\bm{a})$ is expressed
by using the Slater-Koster table \cite{Slater-Koster,Takegahara}
Here we consider only the $f$-electron hopping
through $\sigma$ bond $(ff\sigma)$.

\subsubsection{$f$-$p$ hybridization term}

The $f$-$p$ hybridization term is written as
\begin{equation}
 H_{fp} = \sum_{\bm{k},\mu,\nu,\sigma}\sum_{j=1,2}
 [V^{(j)}_{\bm{k}\mu\nu}
 f^{\dag}_{\bm{k}\mu\sigma} p_{j\bm{k}\nu\sigma}+{\rm h.c.}],
\end{equation}
where $p_{j\bm{k}\mu\sigma}$ is the annihilation operator of
$p$ electron with spin $\sigma$ in the orbital $\nu$ of
$j$-th oxygen and $j$ denotes the label of oxygen ions
in the unit cell, as shown in Fig.~1.
Note that $\nu$ runs among x, y, and z
which correspond to $p_x$, $p_y$, and $p_z$ orbitals, respectively.
The hybridizations $V^{(1)}$ and $V^{(2)}$ are, respectively,
written as
\begin{eqnarray}
  V^{(1)}_{\bm{k}\mu\nu}
  &=&t^{fp}_{\mu\nu}(\bm{b}_1)
  +t^{fp}_{\mu\nu}(\bm{b}_4)e^{-i \bm{k} \cdot \bm{a}_1} \nonumber \\ 
  &+&t^{fp}_{\mu\nu}(\bm{b}_7)e^{-i \bm{k} \cdot \bm{a}_2}
  +t^{fp}_{\mu\nu}(\bm{b}_6)e^{-i \bm{k} \cdot \bm{a}_3}
\end{eqnarray}
and
\begin{eqnarray}
  V^{(2)}_{\bm{k}\mu\nu}
  =t^{fp}_{\mu\nu}(\bm{b}_5)
  +t^{fp}_{\mu\nu}(\bm{b}_8)e^{i \bm{k}\cdot \bm{a}_1} \nonumber \\
  +t^{fp}_{\mu\nu}(\bm{b}_3)e^{i \bm{k}\cdot \bm{a}_2}
  +t^{fp}_{\mu\nu}(\bm{b}_2)e^{i \bm{k}\cdot \bm{a}_3}
\end{eqnarray}
where $t^{fp}_{\mu\nu}(\bm{b})$ denotes the hopping amplitude
between $f$ and $p$ orbitals along $\bm{b}$ direction.
Here we note that $\bm{b}$ runs among
$\bm{b}_1 \sim \bm{b}_8$.
The hopping integral  $t^{fp}_{\mu\nu}(\bm{b})$ is represented
in terms of $(fp\sigma)$ and $(fp\pi)$
by using the Slater-Koster table.\cite{Slater-Koster,Takegahara}

\subsubsection{$p$-electron term}

The $p$-electron part is expressed as
\begin{equation}
  H_{p} = \sum_{\bm{k},\nu,\nu',\sigma}\sum_{i,j}
  [\varepsilon^{(ij)}_{\bm{k}\nu\nu'}+E_p \delta_{ij} \delta_{\nu\nu'}]
  p^{\dag}_{i \bm{k}\nu\sigma}p_{j \bm{k}\nu'\sigma},
\end{equation}
where $\varepsilon^{(ij)}_{\bm{k}\nu\nu'}$ is the $p$-electron dispersion,
$i$ and $j$ denote the label of oxygen ions
in the unit cell, as shown in Fig.1,
and $E_p$ is the $p$-electron level.
Note that we take into account nearest neighbor
and next nearest neighbor hoppings for $p$ electrons.
We also note that
the relations of
$\varepsilon^{(22)}_{\bm{k}\nu\nu'}=\varepsilon^{(11)}_{\bm{k}\nu\nu'}$
and
$\varepsilon^{(21)}_{\bm{k}\nu\nu'}=\varepsilon^{(12)*}_{\bm{k}\nu\nu'}$.

The diagonal part is given by
\begin{eqnarray}
 \varepsilon^{(11)}_{\bm{k}\nu\nu'}
 \!&=&\! 2t_{\nu\nu'}(\bm{a}_1) \cos(k_x/2+k_y/2) \nonumber \\
    \!&+&\! 2t_{\nu\nu'}(\bm{a}_2-\bm{a}_3) \cos(k_x/2-k_y/2) \nonumber \\
 \!&+&\! 2t_{\nu\nu'}(\bm{a}_2) \cos(k_y/2+k_z/2) \nonumber \\
    \!&+&\! 2t_{\nu\nu'}(\bm{a}_3-\bm{a}_1) \cos(k_y/2-k_z/2) \nonumber \\
 \!&+&\! 2t_{\nu\nu'}(\bm{a}_3) \cos(k_z/2+k_x/2) \nonumber \\
    \!&+&\! 2t_{\nu\nu'}(\bm{a}_1-\bm{a}_2) \cos(k_z/2-k_x/2),
\end{eqnarray}
where the hopping amplitudes are given by
\begin{eqnarray}
 t_{\nu\nu'}(\bm{a}_1)&=&
\left(
\begin{array}{ccc}
p_+ & p_- & 0 \\
p_- & p_+ & 0 \\
0   & 0   & (pp\pi)'
\end{array}
\right),\nonumber \\
t_{\nu\nu'}(\bm{a}_2-\bm{a}_3)&=&
\left(
\begin{array}{ccc}
p_+ & -p_- & 0 \\
-p_- & p_+ & 0 \\
0   & 0   & (pp\pi)'
\end{array}
\right),
\nonumber\\
t_{\nu\nu'}(\bm{a}_2)&=&
\left(
\begin{array}{ccc}
(pp\pi)'  & 0  & 0 \\
0  & p_+ & p_- \\
0  & p_- & p_+
\end{array}
\right),
\nonumber \\
t_{\nu\nu'}(\bm{a}_3-\bm{a}_1)&=&
\left(
\begin{array}{ccc}
(pp\pi)'  & 0  & 0 \\
0  & p_+ & -p_- \\
0  & -p_- & p_+
\end{array}
\right),
\nonumber\\
t_{\nu\nu'}(\bm{a}_3)&=&
\left(
\begin{array}{ccc}
p_+  & 0  & p_- \\
0  &  (pp\pi)' & 0 \\
p_-  & 0  & p_+
\end{array}
\right),
\nonumber \\
t_{\nu\nu'}(\bm{a}_1-{\bf a}_2)&=&
\left(
\begin{array}{ccc}
p_+  & 0  & -p_- \\
0  &  (pp\pi)' & 0 \\
-p_-  & 0  & p_+
\end{array}
\right).
\end{eqnarray}
Here $p_{\pm}=[(pp\sigma)' \pm (pp\pi)']/2$,
where $(pp\sigma)'$ and $(pp\pi)'$ denote the Slater-Koster integral
of $p$ electron among next-nearest neighbor oxygen sites.

As for the off-diagonal parts, we obtain
\begin{eqnarray}
 \varepsilon^{(12)}_{\bm{k}{\rm xx}}
 &=&(pp\sigma)[e^{i \bm{k}\cdot(\bm{a}_1+\bm{a}_3)}
           +e^{i\bm{k}\cdot\bm{a}_2}]
\nonumber \\
 &+&
(pp\pi)[e^{i\bm{k}\cdot(\bm{a}_1+\bm{a}_2)}
           +e^{i\bm{k}\cdot\bm{a}_3}]
 \nonumber \\
 &+&
 (pp\pi)[e^{i\bm{k}\cdot(\bm{a}_2+\bm{a}_3)}
           +e^{i\bm{k}\cdot\bm{a}_1}],
\end{eqnarray}
\begin{eqnarray}
 \varepsilon^{(12)}_{\bm{k}{\rm yy}}
 &=&(pp\pi)[e^{i\bm{k}\cdot(\bm{a}_1+\bm{a}_3)}
           +e^{i\bm{k}\cdot\bm{a}_2}]
\nonumber \\
 &+&
(pp\sigma)[e^{i\bm{k}\cdot(\bm{a}_1+\bm{a}_2)}
           +e^{i\bm{k}\cdot\bm{a}_3}]
 \nonumber \\
 &+&
 (pp\pi)[e^{i\bm{k}\cdot(\bm{a}_2+\bm{a}_3)}
           +e^{i\bm{k}\cdot\bm{a}_1}],
\end{eqnarray}
and
\begin{eqnarray}
 \varepsilon^{(12)}_{\bm{k}{\rm zz}}
 &=&(pp\pi)[e^{i\bm{k}\cdot(\bm{a}_1+\bm{a}_3)}
           +e^{i\bm{k}\cdot\bm{a}_2}]
\nonumber \\
 &+&
(pp\pi)[e^{i\bm{k}\cdot(\bm{a}_1+\bm{a}_2)}
           +e^{i\bm{k}\cdot\bm{a}_3}]
 \nonumber \\
 &+&
 (pp\sigma)[e^{i\bm{k}\cdot(\bm{a}_2+\bm{a}_3)}
           +e^{i\bm{k}\cdot\bm{a}_1}].
\end{eqnarray}
Other off-diagonal components are all zeros.

\subsection{Parameters of the model}

The tight-binding Hamiltonian includes many parameters.
Here we try to fix some of them from the experimental and
band-structure calculations results.

{\it (i) CEF parameters.}
It should be noted that it is possible to reproduce the CEF states of
actinide dioxides, when we accommodate plural numbers of $f$ electrons
in the level scheme in which $\Gamma_8$ is lower than $\Gamma_7$.
As already mentioned in Sec.~II,
we obtain $\Gamma_5$ triplet for $n$=2, $\Gamma_8$ quartet for $n$=3,
and $\Gamma_1$ singlet for $n$=4, as experimentally found
in the CEF ground states of UO$_2$ \cite{Amoretti},
NpO$_2$ \cite{Fournier}, and PuO$_2$ \cite{Kern1,Kern2}.
Thus, in the present tight-binding model,
we set $W=-0.01$ eV and $x=0.7$ in order to reproduce
that $\Gamma _8$ quartet is the ground state and $\Gamma _7$ is
the excited state with the excitation energy of about 0.2 eV.

{\it (ii) Spin-orbit coupling.}
From the relativistic band-structure calculation for actinide atom,
the splitting energy between $j$=5/2 and $j$=7/2 states has been found
to be about 1 eV.
Since the splitting energy is given as $(7/2)\lambda$
with the use of spin-orbit coupling $\lambda$,
we fix it as $\lambda=0.3$ eV.

{\it (iii) $f$- and $p$-electron levels.}
In this paper, the $f$-electron level $E_f$ is set as the origin of energy,
leading to $E_f=0$.
On the other hand, the $p$-electron level $E_p$ is considered
to be $E_p=-4$ eV from the comparison of
the relativistic band-structure calculation results.\cite{Maehira}

{\it (iv) Slater-Koster integrals.}
In the model, we use seven Slater-Koster integrals as
$(ff\sigma)$, $(fp\sigma)$, $(fp\pi)$, $(pp\sigma)$, $(pp\pi)$,
$(pp\sigma)'$, and $(pp\pi)'$.
Among them, concerning the $p$-electron hoppings,
we introduce the ratio $\eta$ between nearest and
next nearest neighbor hopping amplitudes,
given by $\eta=(pp\sigma)/(pp\sigma)'=(pp\pi)/(pp\pi)'$.
From the ratio of the distances of nearest and
next nearest neighbor sites,
we set $\eta=(1/\sqrt{2})^{7/2} \approx 0.3$.\cite{Harrison}
As for $(pp\sigma)$ and $(pp\pi)$,
we determine them as $(pp\sigma)$=$0.4$ eV and $(pp\pi)$=$-0.4$ eV,
after several trials to reproduce the structure of the wide $p$ bands
in the relativistic band structure calculations.

Concerning $(ff\sigma)$, it is related with the bandwidth $W$
of $f$ electrons in the $j$=5/2 states on the fcc lattice.
In the limit of infinite $\lambda$,
we have obtained $W$ as
$W=(3/56)(50+2\sqrt{145})(ff\sigma)$$\approx$$4.0(ff\sigma)$.\cite{Hotta-Pu}
Note that for the case of finite $\lambda$,
the width of $j$=5/2 bands is deviated from $W$,
but when $\lambda$ is large enough as in actual actinide compounds,
the bandwidth is found to be almost equal to $W$.
From the comparison with the relativistic band-structure calculation
results, the width of $j$=5/2 bands is 0.5$\sim$0.7 eV,
suggesting that $(ff\sigma)$ is in the order of 0.1 eV.
Then, we set $(ff\sigma)$=0.1 eV in the present model.

In the following calculations, due to the diagonalization of the
Hamiltonian, we depict the tight-binding bands
by changing $(fp\sigma)$ and $(fp\pi)$,
which are believed to be key parameters to understand
the electronic structure of actinide dioxides.

\section{Results}

\begin{figure}[t]
\begin{center}
\includegraphics[width=8cm]{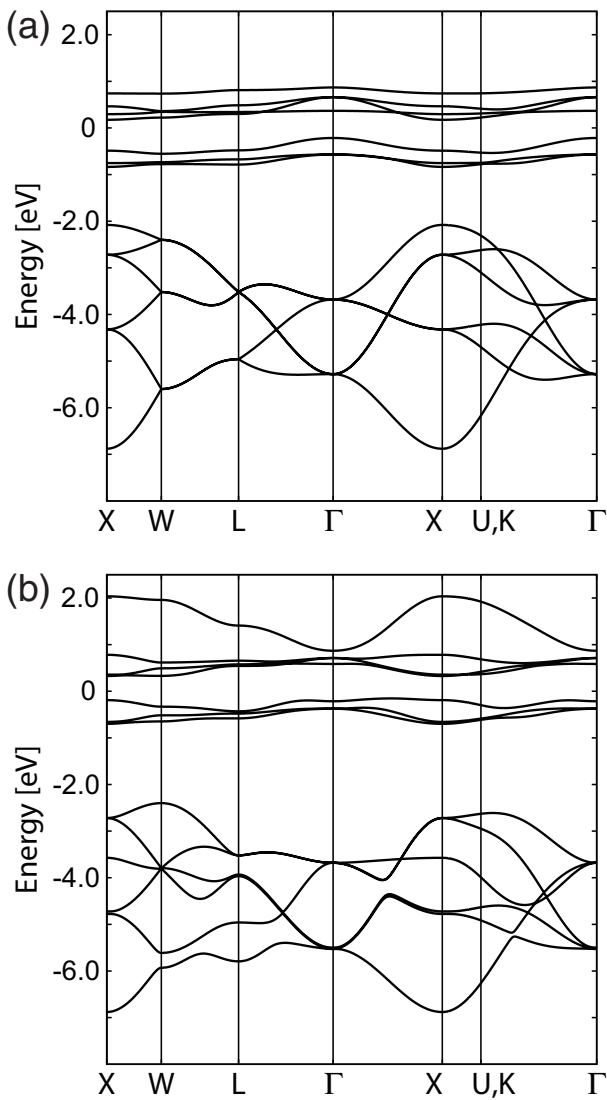}
\caption{Energy band structure obtained by the tight-binding model
for (a) $(fp\sigma)$=$(fp\pi)$=0
and (b) $(fp\sigma)$=$1.0$ eV and $(fp\pi)$=$0.1$ eV.
}
\end{center}
\end{figure}

Now we show our results of the diagonalization of the tight-binding model.
Note that in the following figures of the band structure,
``0'' in the vertical axis indicates the origin of the energy,
not the Fermi level $E_{\rm F}$.
If it is necessary to draw the line of $E_{\rm F}$,
we set it from the condition of $\langle n \rangle$=3 for tetravalent Np ion
in NpO$_2$, where $\langle n \rangle$ denotes the average number of
$f$ electrons per actinide ion.
In the present paper, we do not take care of the difference in actinide ions.

First we consider the case in which the $f$-$p$ hybridization is
simply suppressed.
In Fig.~3(a), we show the tight-binding bands for
$(fp\sigma)$=$(fp\pi)$=0 along the lines in the first Brillouin zone.
We obtain the $f$ and $p$ bands which are not hybridized with each other
and $f$ bands split into $j$=5/2 and $j$=7/2.
Note that $\Gamma_8$ becomes lower than $\Gamma_7$ at the $\Gamma$ point
due to the effect of local CEF potentials.
We observe some degeneracy in $p$ bands
which will be lifted by $f$-$p$ hybridization.

In our first impression, in spite of the simple suppression of
the $f$-$p$ hybridization, 
the overall structure of $f$ and $p$ bands seems to be similar to
that of the relativistic band-structure calculations in Fig.~1.
However, some significant difference is found in the $p$-band structure.
For instance, we find the level crossing in the $p$-band structure
of Fig.~3(a) between the L and $\Gamma$ points,
but we do not observe such behavior in Fig.~1.
Such difference originates from the simplification
to consider only actinide $5f$ and oxygen $2p$ electrons.
The difference in the $p$-band structure is not further discussed
in this paper.

Next we include the $f$-$p$ hybridization as
$(fp\sigma)$=1 eV and $(fp\pi)$=$0.1$ eV in Fig.~3(b).
Due to the effect of $f$-$p$ hybridization,
we find additional dispersion in $f$ and $p$ bands.
In particular, the $p$-band structure becomes similar to
that in the relativistic band-structure calculations.
In this case, we still observe that
$\Gamma_8$ is lower than $\Gamma_7$ at the $\Gamma$ point.

Let us now consider the cases of negative $(fp\pi)$
by keeping the value of $(fp\sigma)$=1 eV.
In Figs.~4(a) and 4(b), we show the results for
$(fp\pi)$=$-0.1$ eV and $-0.6$ eV, respectively.
For $(fp\pi)$=$-0.1$ eV, we do not find significant difference
in the band structure from the case of $(fp\pi)$=$0.1$ eV.
However, for $(fp\pi)$=$-0.6$ eV,
we find that $\Gamma_7$ is lower than $\Gamma_8$
at the $\Gamma$ point.
Regarding the CEF states at the $\Gamma$ point, the $f$-$p$ model with
$(fp\sigma)$=1 eV and $(fp\pi)$=$-0.6$ eV seems to reproduce
the relativistic band-structure calculation results.
Note that in the $p$-band structure, we find the level crossing
of two low-energy bands along the line between W and L points,
which has not been observed in the band-structure calculation.
However, as mentioned above, we do not further pursue
the difference in the $p$-band structure.

\begin{figure}[t]
\begin{center}
\includegraphics[width=8cm]{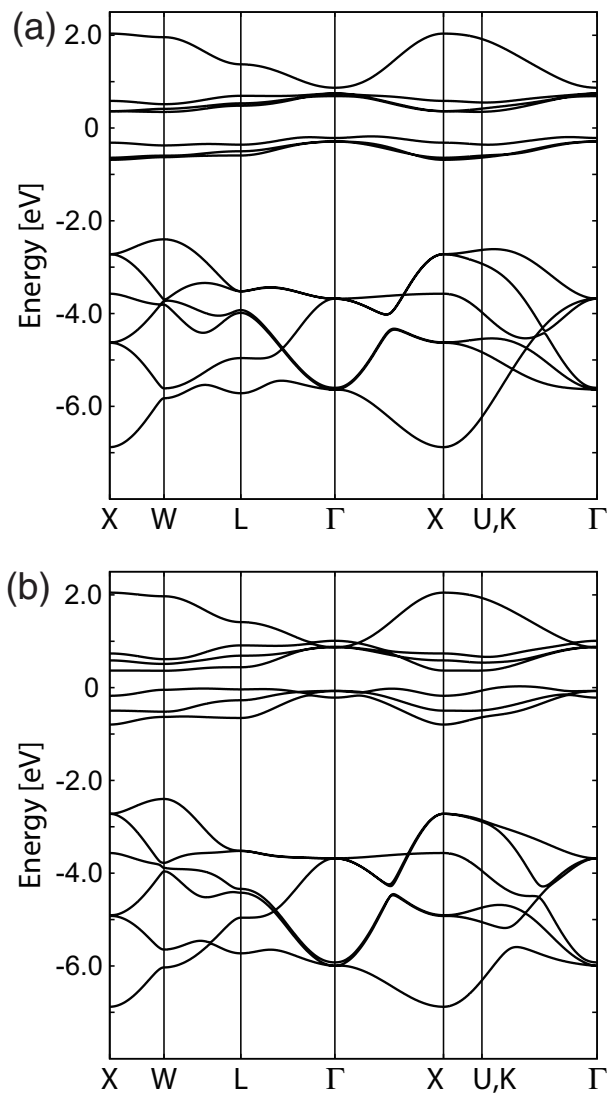}
\caption{Energy band structure obtained by the tight-binding model
for (a) $(fp\sigma)$=$1.0$ eV and $(fp\pi)$=$-0.1$ eV and
(b) $(fp\sigma)$=$1.0$ eV and $(fp\pi)$=$-0.6$ eV.
}
\end{center}
\end{figure}

Here we turn our attention to the $f$-electron states at the $\Gamma$ point.
In the relativistic band-structure calculations
for NpO$_2$,\cite{Maehira}
we have already pointed out that the $\Gamma_7$ level becomes lower than
that of $\Gamma_8$,
in sharp contrast to the local CEF state in the $j$-$j$ coupling
scheme expected from the experimental results.
This is due to the fact that the CEF potential is
not included satisfactorily
in the relativistic band-structure calculation.
On the other hand, the CEF potential is included in the tight-binding model
within the point charge approximation and the change of the level scheme
at the $\Gamma$ point can be explained by the $f$-$p$ hybridization.
When we do not consider the $f$-$p$ hybridization,
we find that $\Gamma_8$ level becomes lower than that of $\Gamma_7$,
but with the increase of the effect of $f$-$p$ hybridization,
the order of the level at the $\Gamma$ point is converted.
Namely, the order of $\Gamma_7$ and $\Gamma_8$ levels is
determined by the competition between the CEF potential
and the $f$-$p$ hybridization.
In this sense, the CEF potential is not included satisfactorily
in comparison with the $f$-$p$ hybridization
in the band-structure calculation.

In the fluorite crystal structure of AnO$_2$,
actinide ion is surrounded by eight oxygen ions in the [111]
and other equivalent directions.
Thus, the $\Gamma_7$ orbital is penalized from the viewpoint of
electrostatic energy, since its wavefunction is elongated
along the [111] directions.
However, the wavefunctions of
two $\Gamma_8$ orbitals are expanded in the directions of axes.
Namely, it is qualitatively understood that $\Gamma_8$ level is lower than
$\Gamma_7$ one in the actinide dioxides.

From the viewpoint of the overlap integral
between actinide $5f$ and oxygen $2p$ electrons,
we expect that the hybridization of
$\Gamma_7$ orbital is larger than that of $\Gamma_8$.
Thus, due to the effect of $f$-$p$ hybridization, the $\Gamma_7$
level becomes lower than $\Gamma_8$, even if the local CEF
ground state is $\Gamma_8$.
When the effect of $f$-$p$ hybridization is relatively larger than
that of the CEF potential,
it is possible to observe that $\Gamma_7$ is lower than $\Gamma_8$,
as actually found in the relativistic band-structure calculation
results.
We emphasize that it is one of the key points concerning the $f$-$p$
hybridization to understand the electronic structure of
actinide dioxides.

\begin{figure}[t]
\begin{center}
\includegraphics[width=8cm]{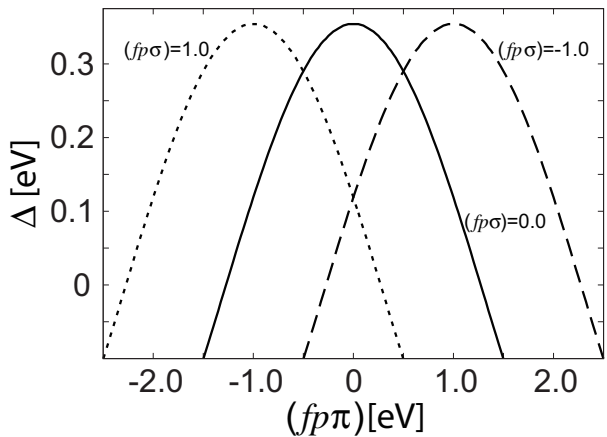}
\caption{Energy difference $\Delta$ between $\Gamma_7$ and $\Gamma_8$ states
at the $\Gamma$ point in the $j$=5/2 bands as a function of $(fp\pi)$ for
$(fp\sigma)$=$0$ eV (solid curve),
$(fp\sigma)$=$1$ eV (broken curve),
and $(fp\sigma)$=$-1$ eV (dotted curve).
A positive $\Delta$ denotes that the energy of $\Gamma_7$ is larger than
that of $\Gamma_8$.
}
\end{center}
\end{figure}

In Fig.~5, we depict the energy difference $\Delta$ between
the $\Gamma_8$ and $\Gamma_7$ states at the $\Gamma$ point
as functions of $(fp\pi)$ for several values of $(fp\sigma)$.
Note that $\Delta$ is positive when $\Gamma_8$ is lower than
$\Gamma_7$.
For $(fp\sigma)$=0, we find that $\Delta$ is positive in the
region of $|(fp\pi)| \alt 1.2$ eV.
When we change the value of $(fp\sigma)$,
$\Delta$ is found to be maximum at $(fp\pi)=-(fp\sigma)$
due to the effective disappearance of the $f$-$p$ hybridization
between actinide $\Gamma_7$ and oxygen $2p$ electrons.

Readers may consider that the absolute value of
$(fp\pi)$ should not be so small only for the purpose
to keep the order of the local CEF states.
However, if we increase the absolute value of $(fp\pi)$ for
$(fp\sigma)=1$ eV, we should remark that the $f$- and $p$-electron
bands are significantly changed from those
in the relativistic band-structure calculation results.
Thus, from the viewpoints of the local CEF states and
the comparison with the band-structure calculations,
the reasonable parameters are found
in the case of small $|(fp\pi)|$ for $(fp\sigma)=1$ eV.

\section{Discussion and Summary}

In this paper, we have analyzed the tight-binding model
for AnO$_2$ in comparison with the local CEF states
and the result of the relativistic
band-structure calculations.
We have concluded that $|(fp\pi)|$ should be small
for the case of $(fp\sigma)=1$ eV in our tight-binding model
in order to keep the CEF levels at the $\Gamma$ point.
We have also emphasized that such a condition coincides with that
for the octupole ordering on the basis of the $f$-$p$ model.\cite{Kubo4}
Namely, the condition to keep the local $\Gamma_8$ ground state is
consistent with the appearance of the ordering of magnetic octupole
which is composed of complex spin and orbital degrees of freedom.

Here we provide a comment on the local CEF state in the
band-structure calculations.
As long as we perform the band-structure calculations
with in the LDA,
it is found  that the $\Gamma_7$ state
becomes lower than the $\Gamma_8$ at the $\Gamma$ point,
in contrast to the local CEF state expected from the experiment.
In this paper, we have proposed the scenario to control the effect
of $f$-$p$ hybridization on the CEF state,
but it should be remarked that in the LDA calculation,
we could $not$ obtain insulating state
corresponding to the multipole ordering for NpO$_2$.\cite{Maehira}
In order to improve this point, we need to consider the effect of
the Coulomb interactions, but it is a serious problem.
One way for this problem is to employ the LDA+$U$ method.
In fact, it has been reported that we the ordered state
with octupole and higher multipoles can be reproduced,\cite{Suzuki}
suggesting that the $\Gamma_8$ state is lower than $\Gamma_7$
in the electronic structure.
The effective inclusion of the Coulomb interaction is
an alternative scenario to understand the CEF state
consistent with the experiments.

Although we have not discussed the difference in electronic structure
due to the change of actinide ions in this paper,
it is naively expected that the difference between $E_f$ and $E_p$
becomes small in the order of Th, U, Np, Pu, Am, and Cm
from the chemical trends in actinide ions and
the previous band-structure calculations.
On the other hand, the change of $f$-$p$ hybridization among actinide
dioxides may play more important role to explain the effect of
the difference in actinide ions.
It is an interesting future problem to clarify the key issue to
understand the difference
in electronic structure of actinide dioxides.

In summary, we have constructed the $f$-$p$ model
in the tight-binding approximation.
We have determined the parameters by the experimental results
and the relativistic band-structure calculations.
It has been concluded that the absolute value of $(fp\pi)$
should be small for $(fp\sigma)$=1 eV
in order to reproduce simultaneously
the local CEF states and the band-structure calculation results.
The small value of $|(fp\pi)|$ is consistent with
the condition to obtain the octupole
ordering in the previous analysis of the $f$-$p$ model.
We believe that the present tight-binding model will useful
to extract the essential point of
the electronic structure of actinide dioxides
from the complicated band-structure calculation results.

\section*{Acknowledgement}

The authors thank S. Kambe, K. Kubo, and Y. Tokunaga
for discussions on actinide dioxides.
This work has been supported by a Grant-in-Aid for
for Scientific Research on Innovative Areas ``Heavy Electrons''
(No. 20102008) of The Ministry of Education, Culture, Sports,
Science, and Technology, Japan.


\end{document}